\documentclass{aastex631}

\usepackage{amssymb}
\usepackage{amsthm}
\usepackage[utf8]{inputenc}
\usepackage{amsmath}
\usepackage{graphicx}
\usepackage{mdwlist}
\usepackage{array}
\usepackage{tabularx}
\usepackage{wasysym}
\usepackage{makecell}
\usepackage{svg}

\begin{document}

\title{Thermal history of the Erg Chech 002 parent body: Early accretion and early differentiation of a small asteroid}

\author[0000-0003-1932-602X]{Wladimir Neumann}
\affiliation{Institute of Geodesy and Geoinformation Science, Technische Universität Berlin \\
Kaiserin-Augusta-Allee 104-106 \\
10553 Berlin, Germany}
\affiliation{Klaus-Tschira-Labor für Kosmochemie, Institut für Geowissenschaften, Universität Heidelberg \\
Im Neuenheimer Feld 234-236 \\
69120 Heidelberg, Germany}
\affiliation{Institute of Planetary Research, German Aerospace Center (DLR) \\
Rutherfordstr. 2 \\
12489 Berlin, Germany}

\author{Robert Luther}
\affiliation{Museum für Naturkunde - Leibniz Institute for Evolution and Biodiversity Science \\
Berlin, Germany}

\author{Mario Trieloff}
\affiliation{Klaus-Tschira-Labor für Kosmochemie, Institut für Geowissenschaften, Universität Heidelberg \\
Im Neuenheimer Feld 234-236 \\
69120 Heidelberg, Germany}

\author{Philip M. Reger}
\affiliation{Department of Earth Sciences, Institute of Earth and Space Exploration, University of Western Ontario \\
N6A 5B7 London \\
Ontario, Canada}

\author{Audrey Bouvier}
\affiliation{Bayerisches Geoinstitut, University of Bayreuth \\
95447 Bayreuth, Germany}

\begin{abstract}

The history of accretion and differentiation processes in the planetesimals is provided by various groups of meteorites. Sampling different parent body layers, they reveal the circumstances of the metal-silicate segregation and the internal structures of the protoplanets. The ungrouped achondrite Erg Chech 002 (EC 002) added to the suite of samples from primitive igneous crusts. Here we present models that utilize thermo-chronological data for EC 002 and fit the accretion time and size of its parent body to these data. The U-corrected Pb-Pb, Al-Mg, and Ar-Ar ages used imply a best-fit planetesimal with a radius of $20-30$ km that formed at $0.1$ Ma after CAIs. Its interior melted early and differentiated by $0.5$ Ma, allowing core and mantle formation with a transient lower mantle magma ocean, and a melt fraction of $<25$ \% at the meteorite layering depth. EC 002 formed from this melt at a depth of $0.8$ km in a partially differentiated region covered by an undifferentiated crust. By simulating collisions with impactors of different sizes and velocities, we analyzed the minimum ejection conditions of EC 002 from its original parent body and the surface composition of the impact site. The magma ocean region distinct from the layering depth of EC 002 implyes that it was not involved in the EC 002 genesis. Our models estimate closure temperatures for the Al-Mg ages as $1060$ K to $1200$ K. A fast parent body cooling attributes the late Ar-Ar age to a local reheating by another, late impact.

\end{abstract}

\keywords{Planetary Interior (1248) --- Planetesimals (1259) --- Asteroids (72) --- Meteorites (1038) --- Achondrites (15) --- Cosmochronology (332) --- Accretion (14) --- Impact Phenomena (779)}

\section{Introduction} \label{intro}

Meteorites sample materials from planetary objects that existed or still exist in the solar system. Non-chondritic space rocks sample differentiated layers of their parent bodies, such as cores, mantles, or crusts. Cosmochemical analyses of various groups of achondrites and primitive achondrites provide a partial chronological picture of accretion and differentiation processes in early solar system's protoplanets. They allow tracing the circumstances of the metal-silicate segregation and the interior structures established thereupon. An early metal-silicate separation that stratified planetesimals and protoplanets at least partially is ascertained by the compositions and chronological records of a variety of differentiated meteorites, for example, HEDs, iron meteorites, grouped achondrites and primitive achondrites, as well as various ungrouped achondrites (e.g., Tafassasset, NWA 011, NWA 6704). Space observations of asteroid belt objects, such as (4) Vesta, complement the meteorite record and provide asteroid-scale case studies for differentiated planetary objects. Structures ranging from undifferentiated objects that experienced incipient melting of chondritic protoliths over partially differentiated objects with chondritic crusts, e.g., (21) Lutetia \cite{Neumann2013}, to fully differentiated bodies with metallic cores, silicate mantles, and silicate crusts, e.g., (4) Vesta \cite{Neumann2014a}, can be derived. Differentiated crustal material of planetesimals has been thought in the past to be represented by basaltic rocks, such as eucrites and some ungrouped achondrites. More recently, discovery of andesitic achondrites suggested that some of the early solar system's planetary objects had andesitic crusts. Andesitic or trachyandesitic achondrites comprise to date NWA 11575, GRA 06128 and GRA 06129, ALM-A, and NWA 11119, and a newly discovered unique achondrite Erg Chech 002 (EC 002) added to the suite of alkaline-rich achondrites that sample primitive igneous crusts\cite{Barrat2021}. Here, we present properties of the parent body and the formation conditions of the EC 002 material as indicated by thermal evolution and differentiation model fits to the chronological record of this meteorite.

\section{EC 002 Properties, Data, and Methods}
Chemical features of EC 002 \cite{Barrat2021} demonstrate that it is a unique crustal fragment of an ancient, differentiated parent body that formed on a similar timescale as the cores of iron meteorite parent bodies\cite{Barrat2021,Kruijer2020}. Its major element chemical composition is distinct from all other basaltic, andesitic, or trachyandesitic achondrites. The parent body of EC 002 can most likely be affiliated to the NC family of objects formed early in the accretion disk, based on the Tm/Tm* ratio identical to those of NC differentiated bodies\cite{Barrat2016} and oxygen isotopic and $^{54}$Cr compositions similar to those of some ungrouped basaltic (NC) achondrites\cite{Barrat2021,Zhu2022}. Alkali concentration and trace element pattern distinguish EC 002 from other andesitic and NC achondrites, indicating, potentially, a distinct parent body\cite{Barrat2021}.
Relatively high abundances of highly incompatible elements, quite high MgO concentration and Mg\#-number, and trace element distribution inconsistent with fractional crystallization indicate that EC 002 is likely a primitive or even a primary melt\cite{Barrat2021}. 
Further analyses by \cite{Barrat2021} showed that such alkali-rich andesitic magmas can be derived from partial melting of chondrites, and those equivalent to EC 002 are obtained after plagioclase exhaustion for a melting degree of $\approx 25$\%, as indicated by the major element composition and consistent with the the unfractionated trace element pattern of EC 002 \cite{Usui2015,Lunning2017}. The MgO content indicates that this magma formed at $1497\pm 20$ K. Ca-rich pyroxene composition indicate a magma crystallization temperature of $1422$ K to $1502$ K with an average of $1459$ K\cite{Barrat2021}, in agreement with typical silicate solidus temperatures inferred for planetesimals. The highest Ca concentration in pyroxene implies a final equilibration temperature of $\approx 1234$ K\cite{Barrat2021}. Interpretations of a fast rock cooling were derived from remnant zoning in groundmass pyroxene and xenocrysts core compositions. Mg\# zoning profile modeling produced a cooling rate of $\approx 5$ °/y between $1473$ and $1273$ K, consistent with a thick lava flow or a shallow intrusion, and a very fast cooling below 1173 K ($>0.1$ to $1$°/d)\cite{Barrat2021}. Such cooling rates could result from an impact that would have excavated or ejected the rock.

Chronological constraints on the thermal evolution of the parent body of EC 002 were derived from analyses of various radioactive decay systems by a few workers. Table \ref{table1} summarizes ages resulting from these analyses along with the respective closure temperatures, if such could be derived from the literature, for the chronometers. Initial $^{26}$Al-$^{26}$Mg systematics \cite{Barrat2021} were based on  analyses of feldspar and pyroxenes, i.e., very high Al/Mg materials. They produced an $^{26}$Al-$^{26}$Mg isochron with the highest value reported for an achondrite of $^{26}$Al/$^{27}$Al$_{\text{initial}}=(5.72\pm 0.07)\times 10^{-6}$ that translates into a closure time of $2.14\pm 0.01$ Myr after CAIs (using an $^{26}$Al half-life of 0.717 Ma by contrast to \cite{Barrat2021}). For an isochron through a bulk, pyroxene, fine-grained, and plagioclase fractions, initial ratios are $(8.89 \pm 0.79) \times 10^{-6}$ (bulk, pyroxene, plagioclase) and $(8.36 \pm 0.49) \times 10^{-6}$ (plagioclase), corresponding to formation times of $1.71\pm 0.11$ Ma and $1.77\pm 0.08$ Ma after CAIs, respectively \cite{Reger2023}. Both values are within the margin of error of each other and older than the Al-Mg age reported by \cite{Barrat2021} by up to half million years. This difference could be on one hand due to potential overcorrection of Mg or miscalibration by using the Miyake-Jima plagioclase standard \cite{Reger2023}. On the other hand, the isochron from \cite{Barrat2021} could represent the closure of pure plagioclase ($^{27}$Al/$^{24}$Mg $> 1500$), while the isochron from \cite{Reger2023} represents the time of closure of a mineral mixture with a higher closure temperature for Mg diffusion. A closure time of $1.71\pm 0.04$ Ma reported by \cite{Fang2022} confirms the Al-Mg result from \cite{Reger2023}. The U isotope composition of EC 002 is heterogeneous, with $^{238}$U/$^{235}$U values of $137.766 \pm 0.027$ for leached pyroxenes and $137.819 \pm 0.007$ for the bulk rock. MC-ICP-MS analysis of the U and Pb isotope compositions of the leached pyroxenes produced a $^{207}$Pb/$^{206}$Pb-$^{204}$Pb/$^{206}$Pb isochron with an age of $4565.87\pm 0.3$ Ma for a $^{238}$U/$^{235}$U ratio of $137.766\pm 0.027$ \cite{Reger2023}. This age corresponds to a closure time of $2.03\pm 0.3$ Ma after CAIs. The $^{207}$Pb-$^{206}$Pb merrillite age was calculated for a $^{238}$U/$^{235}$U ratio of $137.82$ from bulk measurement. The weighted mean of the Pb-Pb ages of seven SIMS analyses of merrillites is $4564.29\pm 2.67$ Ma \cite{Reger2023} and corresponds to a closure time of $3.61\pm 2.67$ Ma.
Both Pb-Pb ages are equal to each other within the margin of error and indicate rapid cooling and absence of significant thermal events after $\approx 3.9$ Ma after CAIs on the parent body \cite{Reger2023}. Furthermore, the Al-Mg model age is consistent within uncertainty with the Pb-Pb age. Using noble gas mass spectrometry and incremental heating, an Ar-Ar plateau age of $4510\pm 40$ Ma and a cosmic ray exposure age of $63.5\pm 24$ Ma were determined \cite{Takenouchi2021}. The Ar-Ar age corresponds to a closure time of $57.9\pm 40$ Ma after CAIs.
\begin{table}
\caption{Time (in Ma rel. to CAIs) and temperature (in K) data used for fitting the parent body. Notes: $^{(\text{fs})}$ feldspar, $^{(\text{px})}$ pyroxene, $^{(\text{b})}$ bulk, $^{(\text{pl})}$ plagioclase, $^{(\text{p})}$ phosphate.
The closure temperatures are from \cite{Goepel1994} (U-Pb-Pb) and \cite{Pellas1997} (Ar-Ar). A CAI age of $4567.9$ Ma was used.
}
\centering
\begin{tabularx}{0.75\columnwidth}{l r r r r c}
\hline \\ [-1.5ex]
Chronometer & \multicolumn{2}{c}{Closure temperature} & \multicolumn{2}{c}{Closure time} & Reference \\
& $T^{c}$ & $\sigma_{T}$ & $t^{c}$ & $\sigma_{t}$ &  \\
& K & K & Ma & Ma & Ma \\ [-1.5ex]
\\ [-1.0ex]
\hline
\\ [-2.0ex]
Al-Mg $^{(\text{fs,px})}$ & - & - & 2.14 & 0.013 & \cite{Barrat2021} \\ 
\\ [-2.0ex]
Al-Mg $^{(\text{b,px,pl})}$ & - & - & 1.83 & 0.12 & \cite{Reger2023, Fang2022} \\ 
\\ [-2.0ex]
Al-Mg $^{(\text{pl})}$ & - & - & 1.9 & 0.09 & \cite{Reger2023} \\ 
\\ [-2.0ex]
U-Pb-Pb $^{(\text{px})}$ & 1085 & 65 & 2.03 & 0.3 & \cite{Reger2023} \\ 
\\ [-2.0ex]
U-Pb-Pb $^{(\text{p})}$ & 770 & 50 & 3.61 & 2.67 & \cite{Reger2023} \\ 
\\ [-2.0ex]
Ar-Ar $^{(\text{pl})}$ & 550 & 20 & 57.9 & 40 & \cite{Takenouchi2021} \\ 
\\ [-2.0ex]
\hline
\end{tabularx}
\label{table1}
\end{table}

While closure temperatures for the Al-Mg system of pyroxene and anorthite have been estimated as $1170$ K to $1270$ K and $970$ K to $1070$ K, respectively \cite{Wadhwa2009}, determining an appropriate closure temperature for the above Al-Mg isochrons is rather difficult, since they were calculated for materials containing fractions of other minerals and would represent time of closure of a mineral mixture. Therefore, we do not specify closure temperatures for Al-Mg ages from \cite{Reger2023,Fang2022} and \cite{Barrat2021}. The pyroxene U-Pb-Pb age at $2.03$ Ma after CAIs corresponds to a closure temperature of $1085$ K, while the younger merrilite U-Pb-Pb age at $6.1$ Ma after CAIs corresponds to a lower closure temperature of $770$ K. In addition, a closure temperature of $550$ K corresponds to a much younger plagioclase Ar-Ar age at $57.9$ Ma after CAIs. Overall, the thermo-chronological data cover a time range of $1.6$ Ma to $98$ Ma after CAIs and a temperature range of $530$ K to $>1090$ K, potentially up to $1300$ K in the case of the Al-Mg chronometer. All these ages post-date the differentiation, since all closure temperatures are lower than the silicate solidus and the upper bound of $1300$ K on the Al-Mg closure temperatures corresponds to only a few percent Fe-FeS melt. The Mg isotopic composition serves as the closest estimate for the timing of the differentiation. The Al-Mg ages of $1.71-1.99$ Ma after CAIs post-date the crystallization of the parent melt. Thus, they post-date the initial melting and differentiation on the parent body, that would start at depth at an even earlier time than at the layering depth of EC 002 and proceed at a higher temperature than those associated with the chronometers. Overall, the data describe the cooling behavior after reaching the $^{26}$Al-induced temperature maximum at the layering depth of Erg Chech 002. The Ar-Ar chronometer that has a much lower closure temperature than other data points provides a possibility to examine whether it reflects a part of the cooling history or a secondary heating and cooling event.
\begin{figure}
\begin{minipage}[ht]{8.5cm}
\setlength{\fboxsep}{0mm}
\centerline{\includegraphics[trim = 10mm 0mm 0mm 0mm, width=8.5cm]{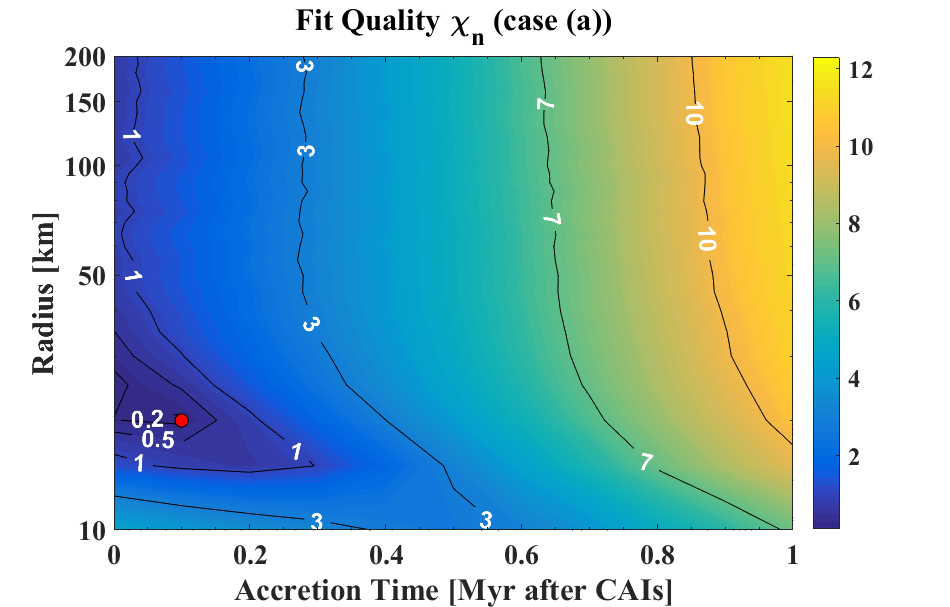}}
\end{minipage}
\begin{minipage}[ht]{8.5cm}
\setlength{\fboxsep}{0mm}
\centerline{\includegraphics[trim = 10mm 0mm 0mm -6mm, width=8.5cm]{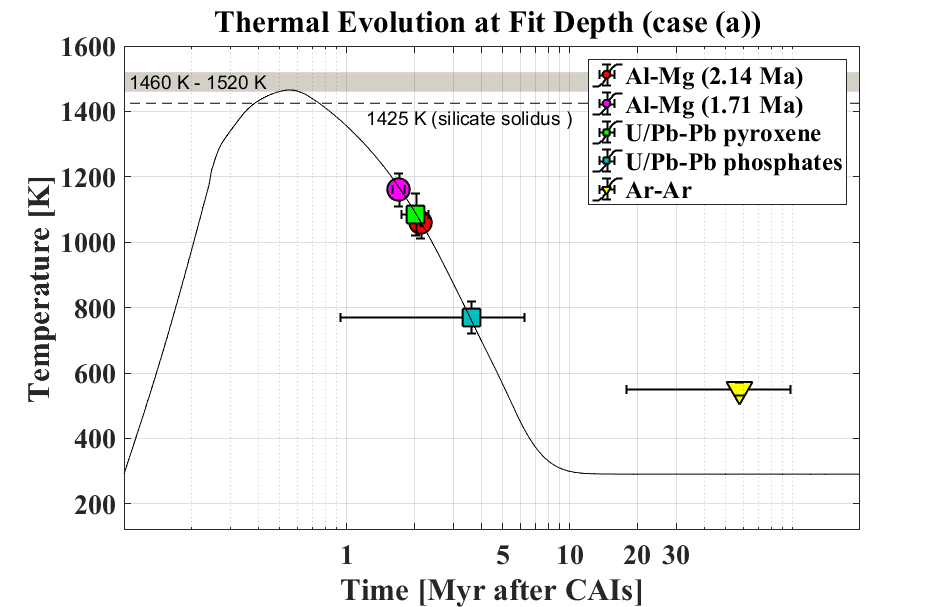}}
\end{minipage}
\\
\begin{minipage}[ht]{8.5cm}
\setlength{\fboxsep}{0mm}
\centerline{\includegraphics[trim = 10mm 0mm 0mm -10mm, width=8.5cm]{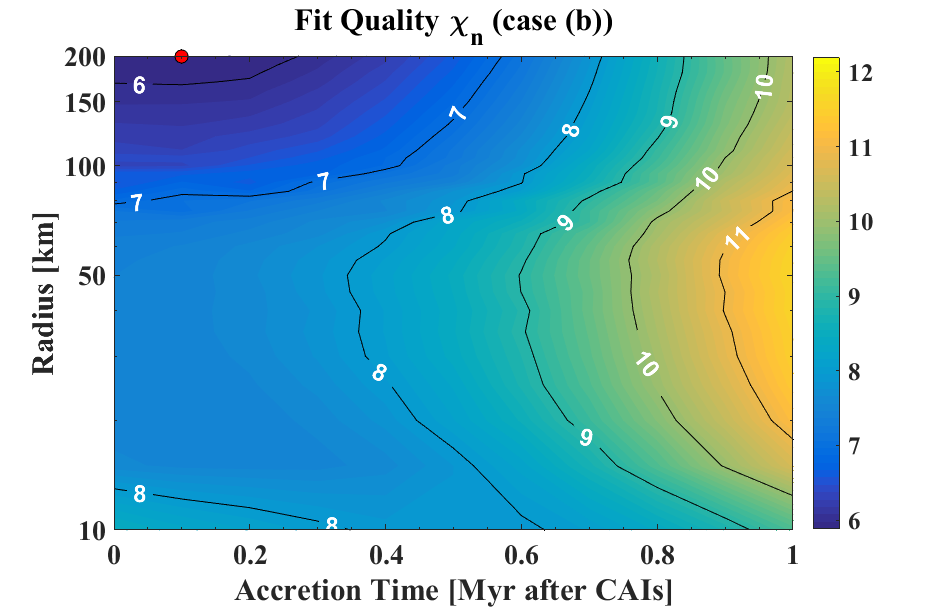}}
\end{minipage}
\begin{minipage}[ht]{8.5cm}
\setlength{\fboxsep}{0mm}
\centerline{\includegraphics[trim = 10mm -7mm 0mm -20mm, width=8.5cm]{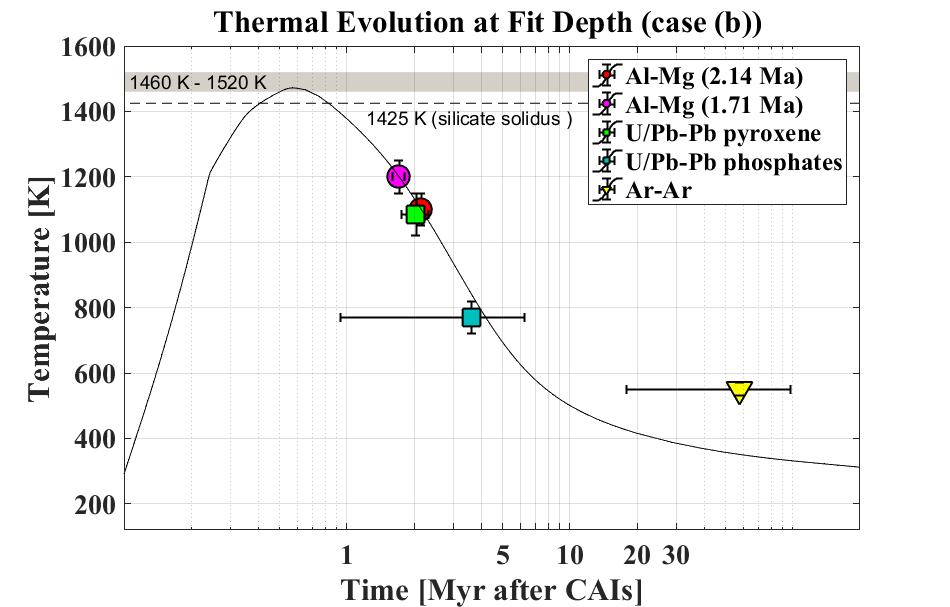}}
\end{minipage}
\caption{Left column: Both colorbar and isolines show fit quality as a function of planetesimal radius $R$ and accretion time $t_{0}$ for EC 002 resulting from a fit of thermo-chronological data without Ar-Ar and Al-Mg ages (top row) or without Al-Mg age but including Ar-Ar age (bottom row). Right column: Thermal evolution at the fit depth of EC 002 for best-fit parent bodies indicated by red dots in the left column.
Best-fit accretion times correspond $<0.2$ Ma (top) and $<0.3$ Ma (bottom). The best-fit radius range is constrained quite well with 20 km to 30 km for the top row, and less well with $>150$ km for the bottom row. The exemplary parent bodies from the respective minimum fit quality regions with $R=20$ km and $t_{0}=0.1$ Ma as well as $R=200$ km and $t_{0}=0.1$ Ma are indicated with red dots.
Both Al-Mg ages are placed directly on the temperature curves. Suggested closure temperatures are $\approx 1160-1180$ K for $t^{c}=1.83$ Ma and $\approx 1060-1100$ K for $t^{c}=2.14$ Ma.}
\label{fig1}
\end{figure}
Since planetesimals are subject to frequent impact events on a range of different scales during their evolution, such an impact could be recorded by the Ar-Ar age. In addition, it could have caused formation of a regolith surface and of simple and complex craters, or even the destruction of the parent body.
We derived parent body properties from the analysis of the thermo-chronological data (Table \ref{table1}) and melting temperature constraints using state-of-the-art global thermal evolution and differentiation models for early solar system planetesimals. In addition, we analysed the minimum impact conditions for the ejection of the meteoroid, which lead to the fall of the EC 002 meteorite, from its formation regime within the original parent body.

The numerical setup is described in detail in the Appendix. The thermal evolution modeling included, in particular, physical processes as Fe,Ni-FeS metal and silicate melting, metal-silicate separation, core and mantle formation, and liquid-state convection in a metal core and in a magma ocean, since melting and (at least partial) differentiation are indicated by Erg Chech 002 composition. Properties of the EC 002 parent body were obtained by approximating the data points with the evolution of the temperature at different depths in the respective time interval using a root mean square (RMS) procedure and by comparing with the magma formation and magma crystallization temperatures\cite{Henke2012,Neumann2018b,Gail2019}. Since closure temperatures associated with the ages are required for a fit procedure, we included only the data points consisting of an age or time after CAIs and a respective temperature interval (Pb-Pb and Ar-Ar ages), but not the Al-Mg ages. Due to a large Ar-Ar age error bar and a weak consistency of this data point with the cooling history, we considered two cases - one where only the U-Pb-Pb data points were fitted and the other with the Ar-Ar data point in addition. An average magma crystallization temperature of $1460$ K and the maximum magma formation temperature of $1520$ K \cite{Barrat2021} were used as an additional constraint to penalize those temperature curves that might fit the data, but have maximum temperatures of either below $1460$ K, or above $1520$ K. This temperature window corresponds to a melt fraction interval of $16$ vol.\% to $32$ vol.\% of a nearly ordinary chondritic protolith. 
The impact simulations were done using the iSALE-2D Eulerian shock physics code \cite{Wuennemann2006} in its Dellen version, which is based on the SALE hydrocode solution algorithm \cite{Amsden1980}. SALE was modified subsequently to include an elastic-plastic constitutive model, fragmentation models, various equations of state, and multiple materials \cite{Melosh1992,Ivanov1997}. Improvements that are more recent include a modified strength model \cite{Collins2004} and the $\varepsilon-\alpha$ porosity compaction model \cite{Wuennemann2006,Collins2011}.
By varying the radius $R$ over a wide range of $10$ km to $200$ km and the accretion time $t_{0}$ between $0$ and $5$ Ma after CAIs for both cases, we constrain a range within the $(R,t_{0})$-diagram appropriate for the parent body of EC 002.

\section{Results}
To demonstrate how the models compare with the meteorite data, we consider the fit quality. Figure \ref{fig1}, left column, shows the normalized fit quality as a function of planetesimal radius $R$ and accretion time $t_{0}$, showing, thereby, two cases - a fit of thermo-chronological data without Ar-Ar and Al-Mg ages denoted as case (a) (top), or without Al-Mg age but including Ar-Ar age, denoted as case (b) (bottom). Best-fit regions are characterized by low values of $\chi_{n}$. Absolute values of the fit quality for objects with the same $R$ and $t_{0}$ depends clearly on the inclusion or exclusion of the Ar-Ar age. A single global minimum is rather difficult to identify for either case, owed to a small number of data points available. However, best-fit objects that are characterized by low $\chi_{n}$ values can be defined. Thus, we characterize a best-fit field by a plateau of the fit quality and an acceptable fit quality by stronger gradients around the boundaries of the plateau. The trends for $\chi_{n}$ are well captured and conclusive constraints on the accretion time and on the parent body size were obtained. A best-fit field for case (a) is confined within $\chi_{n}=0.2$, with parent body radii of $19-27$ km and accretion times of $\leq 0.1$ Ma.
Fits with $\chi_{n}<1$ can be considered as acceptable, with a broader radius range of $15-50$ km and accretion times of $\lesssim 0.3$ Ma. For case (b), a plateau with $\chi_{n}\leq 6$ is obtained for $R\geq 170$ km and $t_{0}\leq 0.27$ Ma, while fits are still considered for $\chi_{n}\leq 7$, with $R\geq 80$ km and $t_{0}\leq 0.57$ Ma. Overall, the result for $t_{0}$ is concordant, with a preferred accretion time of $<0.3$ Ma and formation no later than by $0.6$ Ma. In the terms of the size, the parent body is either small, with a well confined size, for case (a), or very big for a planetesimal, with a preferred lower bound of $170$ km. This is owed to the inclusion of the Ar-Ar age in case (b), since fitting this data point requires a very slow cooling parent body, and this is achieved better for increasingly larger sizes.

Figure \ref{fig1}, right column, shows also the thermal evolution at the layering depth of EC 002 for exemplary best-fit bodies marked with red dots on the $\chi_{n}$ contour plots, those are $R=20$ km, $t_{0}=0.1$ Ma with a layering depth of $0.83$ km for case (a) and $R=200$ km, $t_{0}=0.1$ Ma with a layering depth of $1.26$ km for case (b). By the nature of the fit procedure, the layering depth is the depth at which the minimum value of $\chi_{n}$ was achieved for a given planetesimal, and those values correspond to the data shown on the contour plots. All available data points are shown for the sake of completeness, with the Al-Mg data lacking closure temperatures conveniently shifted onto the curves. The plagioclase Al-Mg data point at $1.9$ Ma from \cite{Reger2023} is omitted since we consider the Al-Mg data point at $1.83$ Ma derived from bulk rock, pyroxene, fine-grained, and plagioclase fractions to be more representative. If the Ar-Ar data is excluded from the procedure (Fig. \ref{fig1}, top right), then both U-Pb-Pb data and the melting temperature window are fitted very well. After the initial heating phase, a maximum temperature of $1466$ K is reached at $0.55$ Ma after CAIs. During the following cooling phase, the U-Pb-Pb data points are crossed at their centers by the temperature curve. The average cooling rate on the nearly linear part of the curve between the peak temperature and $400$ K at $6.6$ Ma after CAIs is $176$ K per $1$ Ma. Consideration of the Ar-Ar age in addition does not result in a good fit of the data (Fig. \ref{fig1}, bottom right). Within a similar general picture, i.e., a temperature maximum of $1472$ K at $0.57$ Ma and a prolonged cooling phase subsequently, the temperature curve is pulled towards the Ar-Ar data point, but without being able to reach its range. As a result, it diverges stronger from the U-Pb-Pb data points compared to case (a). The average cooling rate on the linear part of the curve between the temperature maximum and $600$ K at $6.5$ Ma after CAIs is negligibly slower with $146$ K per $1$ Ma in this case. The absolute value of $\chi_{n}=5.9$ obtained for this body reflects the overall situation. Thus, the Ar-Ar data cannot be explained simply by cooling of the source region, but rather by some secondary event, possibly an impact. Although a best-fit field could be identified for case (b), the comparison of the temperature curve with the Ar-Ar data range clearly demonstrates, that small bodies with low $\chi_{n}$ values derived from case (a) are more likely and preferred candidates for the parent body of EC 002.

\begin{figure}
\begin{minipage}[ht]{4.5cm}
\setlength{\fboxsep}{0mm}
\centerline{{\includegraphics[height=6cm]{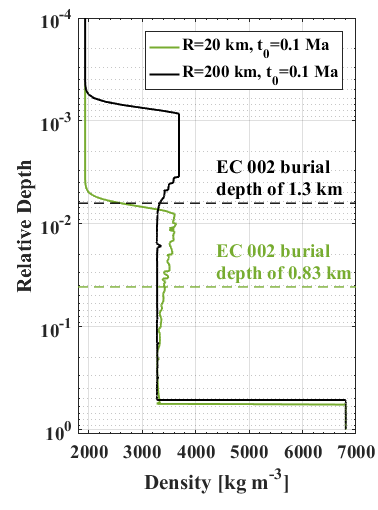}}}
\end{minipage}
\begin{minipage}[ht]{8.5cm}
\setlength{\fboxsep}{0mm}
\centerline{{\includegraphics[width=8.6cm]{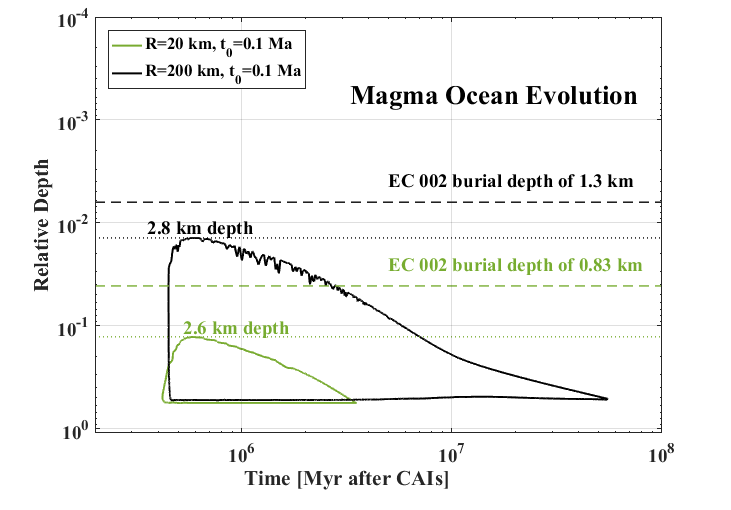}}}
\end{minipage}
\caption{\textit{Left}: Density vs. relative depth profiles after differentiation and cooling (solid lines) and layering depths fitted (dashed lines). \textit{Right}: The extension and evolution of the magma ocean (areas bounded by solid lines) as a function of time and relative depth. layering depths (dashed lines) and upper boundaries of the magma oceans at maximum extension (dotted lines) indicate that melts from a magma ocean likely did not participate in the genesis of the EC 002 material.}
\label{fig2}
\end{figure}

To derive closure temperature estimates for the Al-Mg closure times from \cite{Barrat2021} and \cite{Reger2023}, we consider the temperatures that our best-fit curves produce for these times. Taking a conservative margin of error of $\pm 50$ degrees, the closure temperatures are estimated as $1060\pm50$ K for the Al-Mg age from \cite{Barrat2021} and $1160\pm50$ K for the Al-Mg age from \cite{Reger2023} for case (a) and as $1100\pm50$ K to $1180\pm50$ K for case (b), bracketed overall by $1000$ K and $1230$ K. These intervals agree with lab work derived pyroxene and anorthite closure temperature estimates of $1170$ K to $1270$ K and $970$ K to $1070$ K, respectively \cite{Tourrette1998,Wadhwa2009}.

\begin{figure}
\begin{minipage}[ht]{8.5cm}
\setlength{\fboxsep}{0mm}
\centerline{{\includegraphics[height=6cm]{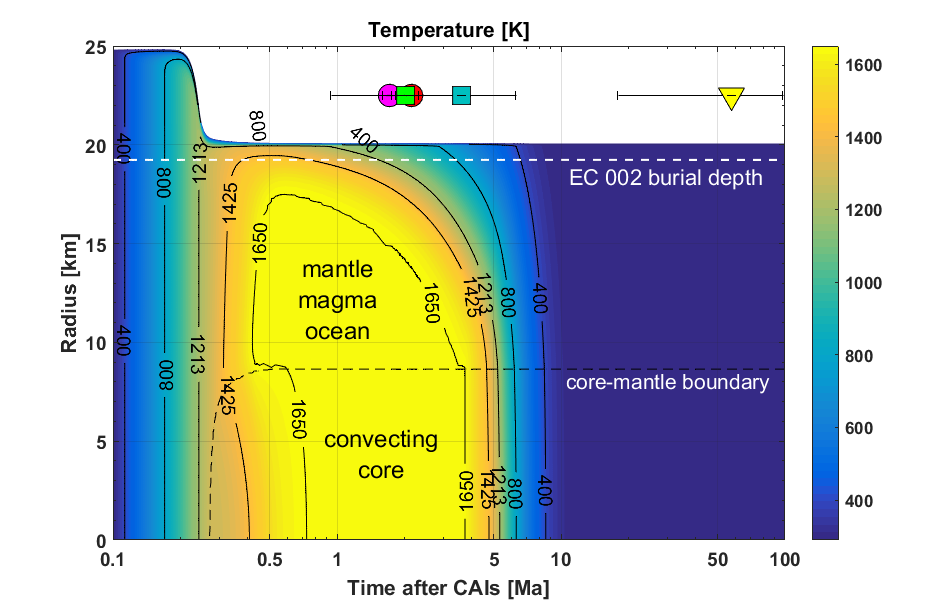}}}
\end{minipage}
\begin{minipage}[ht]{8.5cm}
\setlength{\fboxsep}{0mm}
\centerline{{\includegraphics[height=6cm]{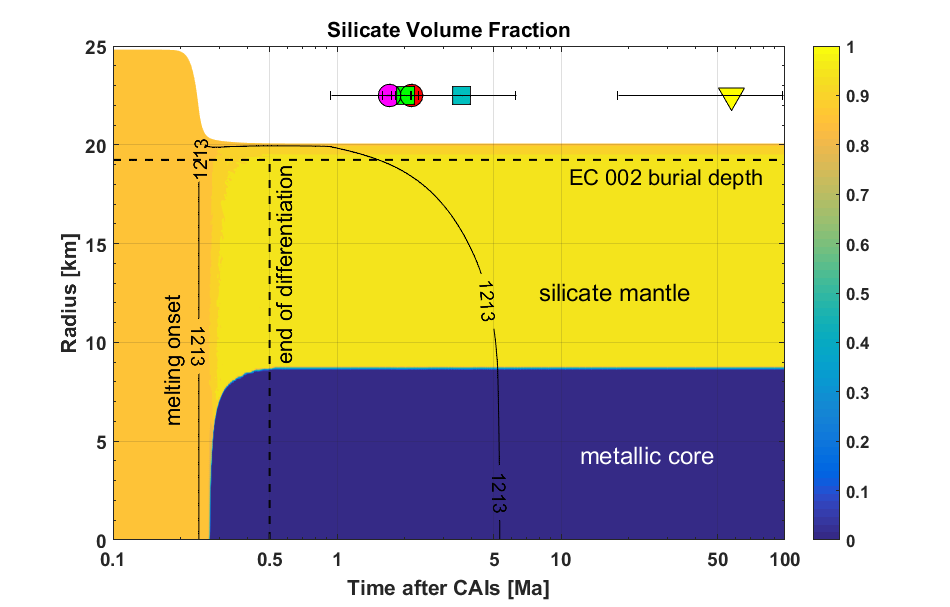}}}
\end{minipage}
\caption{Evolution of the temperature (left panel) and of the silicate volume fraction (right panel) for the preferred best-fit parent body  with a radius of $20$ km and an accretion time of $0.1$ Ma (case (a)). The metal fraction is equal to $1$ minus the silicate fraction. The chronological data points (symbols, cf. Table \ref{table1}), the onset of melting ($1213$ K isoline on right panel), as well as the core-mantle boundary and the layering depth of EC 002 (dashed lines) are shown for comparison.}
\label{fig3}
\end{figure}

Fig. \ref{fig2}, left panel, shows the structure of a parent body via density vs. relative depth profiles. Both small and big parent bodies develop metallic cores, silicate mantles (defined here by a metal volume fraction of $<3$ \%), and partially differentiated layers ($>3$ vol.\% metal) below thin chondritic crusts. The thickness of such a chondritic layer is limited to $0.2-0.7$ km, where the upper $<150$ m remain highly porous and act to slow down cooling of the interior. For a small parent body, the fitted layering depths of EC 002 is located in the partially differentiated layer with $< 4$ vol.\% metal. For a large parent body, it is located in the upper part of the differentiated mantle with $<2$ vol.\% metal. As expected for an accretion close to the formation of the CAIs, the early evolution of both small and large parent body cases is dominated by the heating provided by $^{26}$Al and not by the planetesimal size. In both cases, melting starts early, with the temperature exceeding the metal solidus at $0.24$ Ma and the silicate solidus at $0.3$ Ma at depth (Fig. \ref{fig3}, left panel). The differentiation is finished by $0.5$ Ma (Fig. \ref{fig3}, right panel), though a higher fraction of the interior is differentiated in case (b). The differentiated mantles of both bodies develop turbulently convecting, yet progressively cooling and shrinking magma oceans. The magma ocean onset time is $0.42$ Ma for case (a) and slightly later at $0.448$ Ma for case (b) (Fig. \ref{fig2}, right panel) owed to a more efficient compaction of a $R=200$ km body and a weaker insulation by a thinner porous chondritic crust. The magma ocean life time is $3$ Ma (small parent body) or $55$ Ma (large parent body) until a complete top-down solidification. The one order of magnitude difference between the magma ocean life times stems from a weaker surface cooling of a larger object with $R=200$ km resulting from a smaller surface to volume ratio. layering depths derived for EC 002 and upper boundaries of the magma oceans at their maximum extensions indicate that magma ocean melts likely did not participate in the genesis of the EC 002 material. Overall, the layering depth obtained for EC 002 ranges between $\approx 0.8$ km and $\approx 1.3$ km (objects with radii of $20$ to $30$ km, case (a)) and $\approx 1.2$ km and $\approx 1.3$ km (objects with radii of $180$ to $200$ km, case (b)).
According to our modeling, the parent magma of EC 002 forms \textit{in situ} by partial metal and silicate melting of a nearly chondritic precursor. In case (a), the metal separates from the silicates at the best-fit depth between $0.28$ Ma and $0.5$ Ma, such that a fraction of metal sinks from the source region to the core and is substituted by some silicate melt that migrated from below to EC 002 source depth. The source region keeps a metal fraction of $<4$ vol.\%. After a temperature maximum at $0.55$ Ma, the silicate melt crystallization occurs at $0.75$ Ma and the metal melt crystallization at $1.533$ Ma. For case (b), the metal-silicate separation at EC 002 source depth is quasi-instantaneous due to a higher percolation velocity owed to a higher gravity at $R=200$ km. It occurs within $15$ thousand years between $0.275$ Ma and $0.29$ Ma. The temperature maximum time of $0.57$ Ma is similar to case (a), while both silicates and leftover $<2$ vol.\% metal crystallize slightly later at $0.825$ Ma and $1.658$ Ma, respectively.
\begin{deluxetable}{c c c c c c c}
\label{table2}
\tabletypesize{\scriptsize}
\tablewidth{0.85\columnwidth}
\tablecaption{Impact model parameters overview.}
\centering
\tablehead{ & \makecell{Projectile\\diameter} & \makecell{Cells per\\projectile \\radius} & \makecell{Cell\\length} & \makecell{Crust\\thickness} & \makecell{Impact\\velocity} &\\
& $L$ & - & $dx$ & - & $v_{\text{Imp}}$ \\
& m & - & - & cells & m s$^{-1}$ }
\startdata
& 1000 & 20 & 25 & 5 & 500 \\
\\ [-2.0ex]
& 2000 & 40 & 25 & 5 & 500 \\ 
\\ [-2.0ex]
& 3000 & 60 & 25 & 5 & 500 \\
\\ [-2.0ex]
& 4000 & 80 & 25 & 5 & 500 \\
\\ [-2.0ex]
& 5000 & 100 & 25 & 5 & 500 \\
\\ [-2.0ex]
& 500 & 10 & 25 & 5 & 5000 \\ 
\\ [-2.0ex]
& 700 & 14 & 25 & 5 & 5000 \\
\\ [-2.0ex]
& 1200 & 24 & 25 & 5 & 5000 \\
\\ [-2.0ex]
& 1600 & 32 & 25 & 5 & 5000 \\
\\ [-2.0ex]
& 2000 & 40 & 25 & 5 & 5000 \\
\hline
\hline
& \multicolumn{5}{c}{Material models (strength)} & \\
& \multicolumn{3}{c}{Intact material} & \multicolumn{3}{c}{Damaged material}\\
\hline \\ [-1.5ex]
Layer & Cohesion & \makecell{Coefficient\\of friction} & \makecell{Hugoniot\\elastic limit} & Cohesion & \makecell{Coefficient\\of friction} & \makecell{Hugoniot\\elastic limit} \\
 & $Y_{0,i}$ & $f_{i}$ & $Y_{max,i}$ & $Y_{0,d}$ & $f_{d}$ & $Y_{max,d}$ \\
 & Pa & - & Pa & Pa & - & Pa \\
\hline
Crust & - & - & - & 1.4e3 & 0.77 & 1e9 \\
Mantle & 1e7 & 1.2 & 3.5e9  & 1.0e4 & 0.6 & 3.5e9 \\
Core & 1e8 & - & - & - & - & - \\
\hline
\hline
& \multicolumn{5}{c}{Material models (porosity)} & \\
\hline \\ [-1.5ex]
\makecell{Initial\\porosity} & \multicolumn{2}{c}{\makecell{Initial\\distension}} & \makecell{Elastic\\volumetric\\threshold} & \makecell{Compaction\\efficiency} & \makecell{Sound\\speed\\ratio} & \makecell{Transition\\distension} \\
$\Phi$ & \multicolumn{2}{c}{$\alpha_{0}$} & $\varepsilon_{e}$ & $\kappa$ & $\chi$ & $\alpha_{x}$ \\
\% & \multicolumn{2}{c}{-} & - & - & - & - \\ [-1.5ex]
\\ [-1.0ex]
\hline
\\ [-2.0ex]
41.9 & \multicolumn{2}{c}{$1.720207$} & -2e8 & 0.96 & 0.3 & 1.1 
\enddata
\end{deluxetable}

In general, accretion at optimized times of $\lesssim 0.3$ Ma and a resulting early differentiation cause a transiently inverted temperature profile with a peak in the mantle due to the concentration of $^{26}$Al and a colder core, in agreement with previous planetesimal models \cite{Neumann2012}. If compared over the entire thermal evolution, the best-fit models have overall peak temperatures of $1658$ K attained at the center in case (a), or of $1662$ K attained in the mid-mantle and exceeding the central peak temperature by $\approx 10$ degrees in case (b). On the global scale these maximum temperatures imply an extensive melting in the interior of the parent body (Fig. \ref{fig3}, left panel). In addition to the convecting magma ocean in the mantle, both the melt fraction in the metallic core and the heat flux at the core-mantle boundary are sufficiently high for thermal convection in the core that lasts for only $<4$ Ma in case (a) (Fig. \ref{fig3}, left panel) and up to $80$ Ma in case (b). As with the magma ocean duration, this contrast stems from a slower cooling of a $R=200$ km body. Locally, the temperature evolution at the best-fit depth agrees with the petrologic constraints \cite{Barrat2021,Reger2023}, while only a small leftover metal fraction of $<4$ vol.\% (Fig. \ref{fig3}, right panel) agrees with rare troilite, Fe-Ni metal and iron oxides present in the EC 002 composition. Establishing the temperature fit to different thermo-chronological data points at one and the same depth below a thin chondritic crust inside the parent body supports an \textit{in situ} formation and an intrusive origin for EC 002.
\begin{figure}
\setlength{\fboxsep}{0mm}
\centerline{\includegraphics[trim = 0mm 0mm 0mm 0mm, clip, height=10cm]{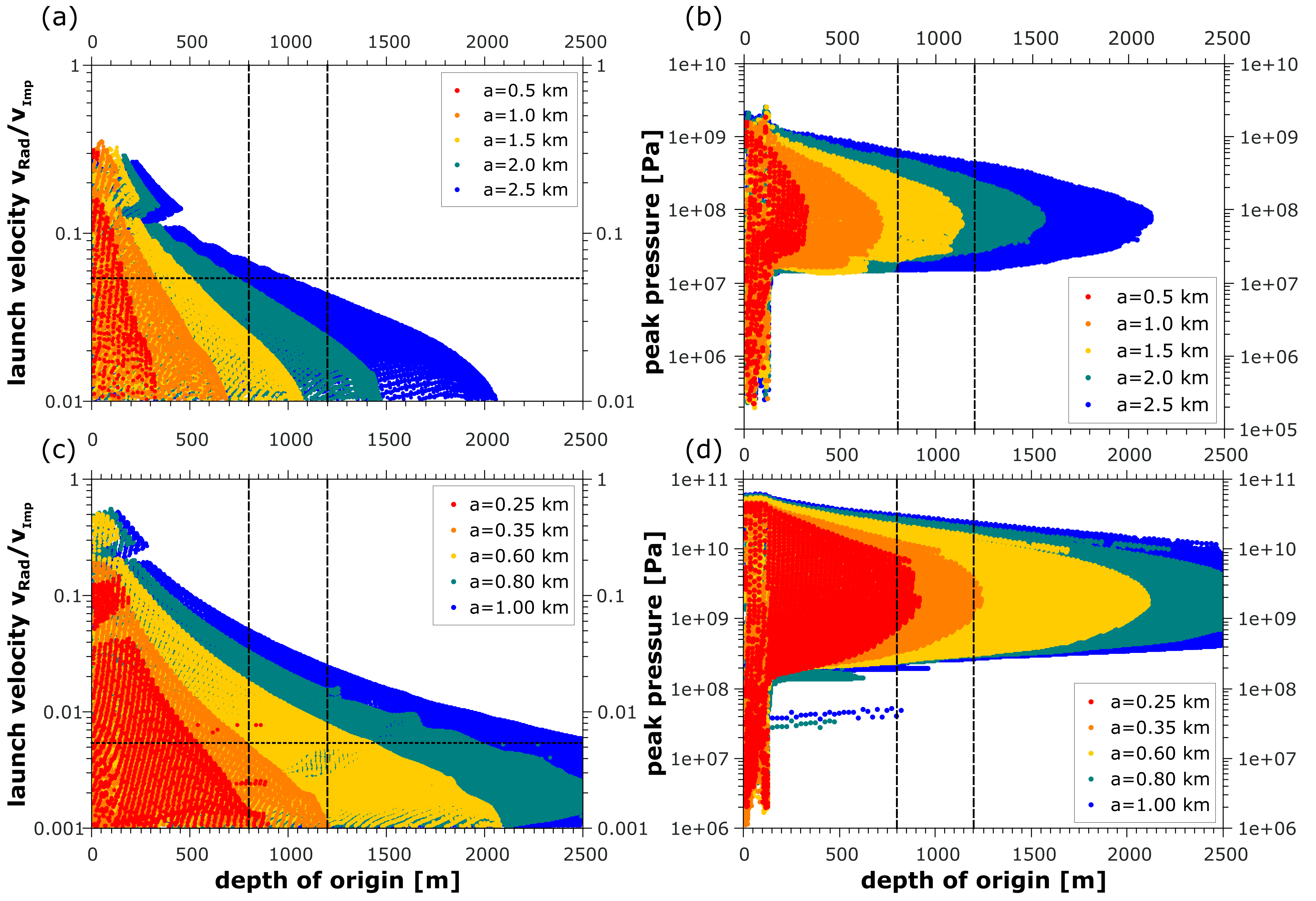}}
\caption{Ejected Material. The launch velocities in radial direction (a,c) and peak pressure (b,d) by which the material was affected are plotted against the depth from where the material originates from for the two impact velocities of 500 m s$^{-1}$ (a,b) and 5000 m s$^{-1}$ (c,d). Black dashed vertical lines indicate the depth of excavation to reach the formation depth of EC 002. Horizontal dotted lines indicate the escape velocity of the body. Colors correspond to the different projectile radii.}
\label{fig31}
\end{figure}

To analyse the minimum ejection conditions of EC 002 from its original main body, we simulated impacts of different sized bodies (Table \ref{table2}) with our $20$ km preferred parent body as target object at a typical main belt impact velocity of $5$ km s$^{-1}$ \cite{Farinella1992}. To analyse velocity related effects, we also simulated a series of impacts with $10$ times smaller impact velocity, which is similar to "slow merger" scenarios. The target is set up as a $3$-layered spherical object according to the structure shown in Fig. \ref{fig2}, left panel. The metallic core, for which we use a Tillotson equation of state for iron and a constant Von-Mises yield strength of $100$ MPa (Table \ref{table2}), extends over $8.73$ km and is resolved by $349$ cells per radius. To simulate the mantle material, we use similar parameters as \cite{Collinson1994}, applying a rheology model with intact and damaged yield surface (Rock model, Table \ref{table2}) and an ANEOS equation of state for dunite. The mantle extends close to the surface and is covered by a $\approx 120$ m thick crust. The crust, which is approximated with 5 cells per radius, is assumed to show regolith-like characteristics. Consequently, we apply a Drucker-Prager rheology model (Table \ref{table2}), the ANEOS for dunite, and the $\varepsilon-\alpha$ porosity compaction model with a porosity of $42$ \% (Table \ref{table2}), using parameters for regolith simulant, which has been used in laboratory impact experiments and has been applied in numerical simulations before \cite{Chourey2020,Luther2022}. The surface gravity of the radial field is $\approx 0.02$ m s$^{-2}$ and the escape velocity is $\approx 27$ m s$^{-1}$. For the projectiles, which are between $500$ m and $5$ km in diameter, we use a similar material model as for the mantle. 
To reduce the computational demands, we only simulate the upper half of the target, neglecting any effects on the opposing hemisphere. The grid consists of $800$ times $900$ cells in horizontal and vertical direction in the high-resolution zone, and $50$ cells of slowly increasing grid size are added to the bottom, top and right side of the high resolution zone. The model applies a cylindrical symmetry so that the left boundary aligns with the symmetry axis of the target. 
The material ejection is based on the analysis of Lagrangian tracer particles, which move following the material flow. We apply the ejection criterion as described by \cite{Luther2018} and the modifications to curved targets shown by \cite{Gueldemeister2022}, and use an ejection altitude of $10$ cells. For details and validation against laboratory experiments, see \cite{Luther2018,Luther2022} and references therein.

To determine the minimum ejection conditions, the ejected material needs to origin from the predicted depth of formation within the target and its ejection velocity is requested to exceed  the escape velocity of $\approx 27$ m s$^{-1}$ of the system. The simulation results are shown in Figure \ref{fig31} for both impact velocities, 500 m s$^{-1}$ (Fig. \ref{fig31}a) and 5 km s$^{-1}$ (Fig. \ref{fig31}c). For a 2D visualization of the initial position of ejected projectile and target material see the Appendix. For the case of 500 m s$^{-1}$ impact velocity, projectile radii of 0.5 km and 1 km do not reach the required excavation depth. Projectiles with 1.5 km and 2 km radii are sufficient to excavate from the formation depth of EC 002, however the ejection velocities are too slow to allow the material to escape from the body. A 4-5 km diameter projectile is sufficient to excavate material from the upper boundary of the formation region, and to eject some of it with sufficient velocity to leave the gravitational system of the parental body. For the case of 5000 m s$^{-1}$ impact velocity, a projectile radius of 0.25 km barely excavates from close to the upper boundary of the formation depth of the EC 002 material. For a radius of 0.35 km, the fastest material from that depth is excavated with a velocity just below the escape velocity of the system.

The modification level of the ejected material due to the impact can be estimated based on the peak pressures to which the material was exposed to (see Figure \ref{fig31}b and d). Light modifications like planar fractures occur for shock pressures of some GPa, while, e.g., whole rock melting of the target requires larger pressures of $\approx 50$ GPa. For an impact velocity of 500 m s$^{-1}$, peak pressures for material excavated from the relevant depth are between 10 MPa and 1 GPa. For 5000 m s$^{-1}$, these pressures range from 200 MPa to $\approx 20-40$ GPa.
\begin{figure}
\begin{minipage}[ht]{8.5cm}
\setlength{\fboxsep}{0mm}
\centerline{{\includegraphics[height=6cm]{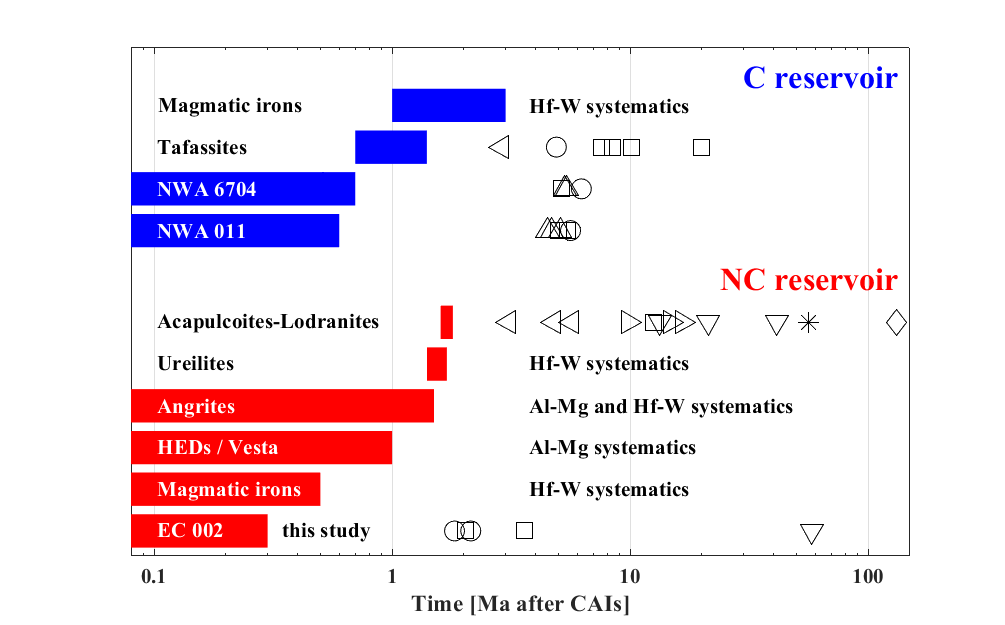}}}
\end{minipage}
\caption{Accretion times of parent bodies of some achondritic and primitive achondritic meteorites (colored patches) derived from model fits to meteorite chronology (data points and text statements). The accretion times and thermo-chronological data are from this study (EC 002), \cite{Kleine2020} and \cite{Neumann2018a} (magmatic iron meteorites), \cite{Neumann2014a} (HEDs / Vesta), \cite{Budde2015} (ureilites), \cite{Kleine2012} and \cite{Schiller2016} (angrites), \cite{Neumann2018b} (acapulcoites-lodranites), \cite{Neumann2023} (NWA 011 and NWA 6704), and \cite{Ma2022} (tafassites). The data points are derived from following chronometers: Hf-W (left-pointing triangles), Mn-Cr (circles), U-Pb-Pb (squares), Al-Mg (up-pointing triangles), I-Xe (right-pointing triangles), Ar-Ar (down-pointing triangles), U-Th-He (star), and Pu fission tracks (diamond). For the data, see Table \ref{table1} (EC 002) and papers cited (all other meteorites).}
\label{fig4}
\end{figure}

\section{Discussion and Conclusions}
Our model fits constrained the EC 002 parent body accretion time and size. The results are conclusive with respect to the accretion time that is $\lesssim 0.3$ Ma after CAIs for acceptable fits and $\leq 0.1$ Ma for best fits. With respect to the parent body size, we obtain either small bodies radii of $20-30$ km, or large ones with radii of $\geq 170$ km, with a preference of a smaller parent body.
This accretion time is very close to those derived for several classes of early solar system's objects that have experienced high-temperature metamorphism, extensive melting, and metal-silicate differentiation, e.g., parent bodies of HEDs (Vesta) \cite{Neumann2014a}, magmatic iron meteorites \cite{Kruijer2020,Neumann2018a}, angrites \cite{Kleine2012,Schiller2016}, and C achondrite grouplets NWA 011 and NWA 6704 \cite{Neumann2023}. However, it pre-dates the accretion of objects that experienced a low-degree melting and only partial metal-silicate differentiation, e.g., parent bodies of primitive achondrites acapulcoites-lodranites \cite{Neumann2018b}, ureilites \cite{Budde2015}, and tafassites \cite{Ma2022}. The accretion times are compared in Fig. \ref{fig4}. While an early accretion of the EC 002 parent body contradicts seemingly a parent melt fraction that is rather similar to the melting degrees of primitive achondrites, the contradiction is resolved by a very shallow layering depth of EC 002 within a fast-cooling small parent object that produced a high degree of melting in its deeper interior.
\begin{figure}
\begin{minipage}[ht]{8.5cm}
\setlength{\fboxsep}{0mm}
\centerline{{\includegraphics[height=6cm]{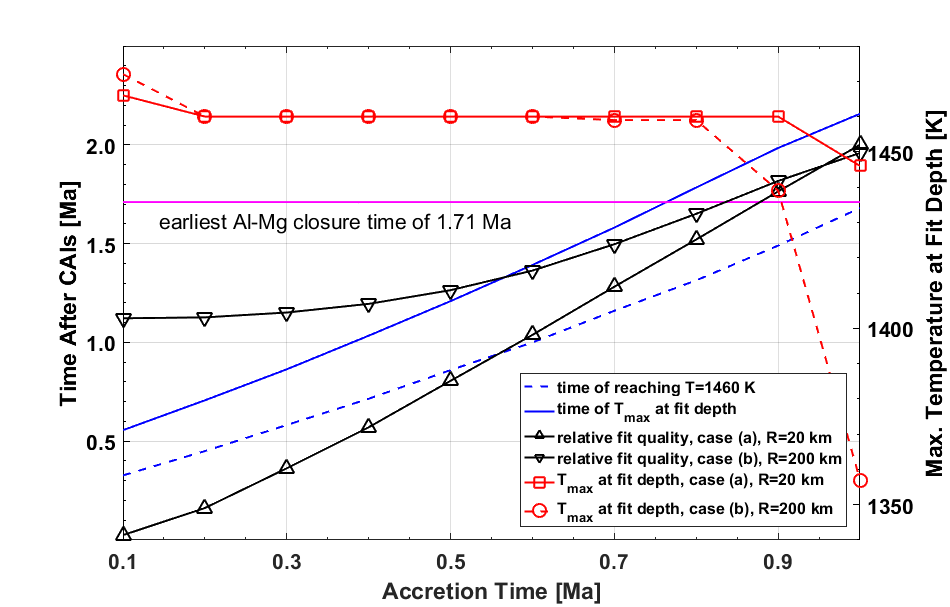}}}
\end{minipage}
\caption{Times of reaching selected temperatures (blues lines), relative normalized fit qualities (black lines), and maximum temperatures at fit depths (red lines) as functions of the accretion time. Two exemplary bodies for Fig. \ref{fig1} with $R=20$ km and $R=200$ km are considered. For these two bodies, the times at which $T=1460$ K is reached for the first time and the times at which the maximum temperature is reached at the fit depth are within $50$ thousand years from each other (thus, blue lines stand for both radii). The relative normalized fit quality is shown for fixed radii and varying accretion time and calculated from $\chi_{n}$ (Fig. \ref{fig1}) by dividing by the maximum value of of $10.5$ obtained within $1$ Ma after CAIs. It varies between $0$ and $1$, but the figure does not provide a scaling. The maximum temperatures at fit depth are similar to identical for early accretion but distinct after $t_{0}=0.9$ Ma.}
\label{fig5}
\end{figure}

The size of the EC 002 parent body of $R=20$ km is close to the lower end of the size distribution for planetesimals. However, such objects were more common in the early asteroid belt than those with $R\approx 200$ km \cite{Bottke2005a}. On the other hand, they would be able to disrupt on a timescale of a few ten Ma \cite{Bottke2005b}, similar to the Ar-Ar age of EC 002.
Since the case (b) best-fit temperature curve (Fig. \ref{fig1}, bottom right panel) was actually not even closely able to fit the Ar-Ar data point, not even within the margin of the closure time error of $\pm 40$ Ma, the Ar-Ar age likely reflects a secondary event, such as a late impact at $t\geq 20$ Ma. This implies that the meteorite was not ejected before $20$ Ma, since it was able to record this impact, contradicting a rapid ejection suggested by \cite{Barrat2021}. In fact, the Ar-Ar age itself \cite{Takenouchi2021} contradicts an earlier ejection. We estimated, further, the Al-Mg closure temperatures as $1160$ K or $1180$ K for the bulk rock, plagioclase, fine-grained, and pyroxene fractions \cite{Reger2023} and $1060$ K or $1160$ K for feldspar and plagioclase fractions \cite{Barrat2021}.
\\
Our melting and metal-silicate separation results show that chondritic material could be preserved on the EC 002 parent body, but only in a thin $<0.2-0.7$ km surface layer. However, preservation of such undifferentiated material is not required to reproduce EC 002 composition, since this meteorite is not chondritic. Nor is a chondritic crust necessary to reproduce EC 002 age systematics, since a largely differentiated layer resulting from partial melting of a nearly chondritic protolith serves as the source layer where EC 002 formed \textit{in situ} and was excavated by an impact. This is supported by uniform pyroxene compositions that indicate that Erg Chech
002 likely represents close to an \textit{in situ} crystallized melt and not a cumulate \cite{Nicklas2022}.
\\
Having suggested a fast cooling, \cite{Barrat2021} inferred from it that the Al-Mg age can be considered to date the crystallization of the parent melt and suggested a differentiation time of $0.95-2.2$ Ma. In contrast, our models imply an earlier differentiation of the parent body $<0.5$ Ma after CAIs (Fig. \ref{fig3}). We note that the differentiation time interval from \cite{Barrat2021} would contradict the Al-Mg model age from \cite{Fang2022} and \cite{Reger2023} that post-date the differentiation of the EC 002 source melt.

The Al-Mg age from \cite{Barrat2021} as such dates the closure temperature of Mg diffusion in plagioclase and not the melt crystallization that pre-dates the closure temperature. Therefore, the crystallization occurred moderately to substantially earlier than the Al-Mg age, depending on the rate of cooling. Our calculations suggest, on one hand, a slow cooling. On the other hand, best-fit temperature curves fall below the silicate solidus of $1425$ K case-dependent at $0.75$ Ma or $0.82$ Ma, implying that the melt crystallized up to $1.4$ Ma before the plagioclase Al-Mg closure time.
As shown in Fig. \ref{fig1}, right panels, and Fig. \ref{fig3}, left panel, for an early accretion with strong $^{26}$Al heating, a time of $\approx 0.4-0.5$ Ma is required for the interior at different depths to heat to the respective maximum temperature and to produce the source melt. Therefore, even assuming a rapid cooling within a few decades after the source material formation, as suggested by \cite{Barrat2021}, the accretion of the parent body must have occurred early. A statement that is detached from any fit procedure is the earliest time at which a minimum temperature of $1460$ K required for the production of the EC 002 parent melt is reached at any depth (Fig. \ref{fig5}, dashed blue line). A comparison of these times for different accretion times with the earliest Al-Mg closure time of $1.83$ Ma (Fig. \ref{fig5}, pink line) shows that the parent body must have accreted not later than $1$ Ma after CAIs. A similar fit-dependent comparison for reaching a maximum temperature at the respective fit depth shows that the parent body needs to accrete before $0.8$ Ma after CAIs even when a rapid cooling is assumed. In addition, far too low maximum temperatures obtained at fit depths (red lines) exclude accretion times of $> 0.8$ in both case (a) and (b). Finally, relative $\chi_{n}$ curves (black lines) confirm the result from Fig. \ref{fig1} that prefers a very early accretion.
\begin{figure}
\setlength{\fboxsep}{0mm}
\centerline{\includegraphics[trim = 0mm 0mm 0mm 0mm, clip, height=7cm]{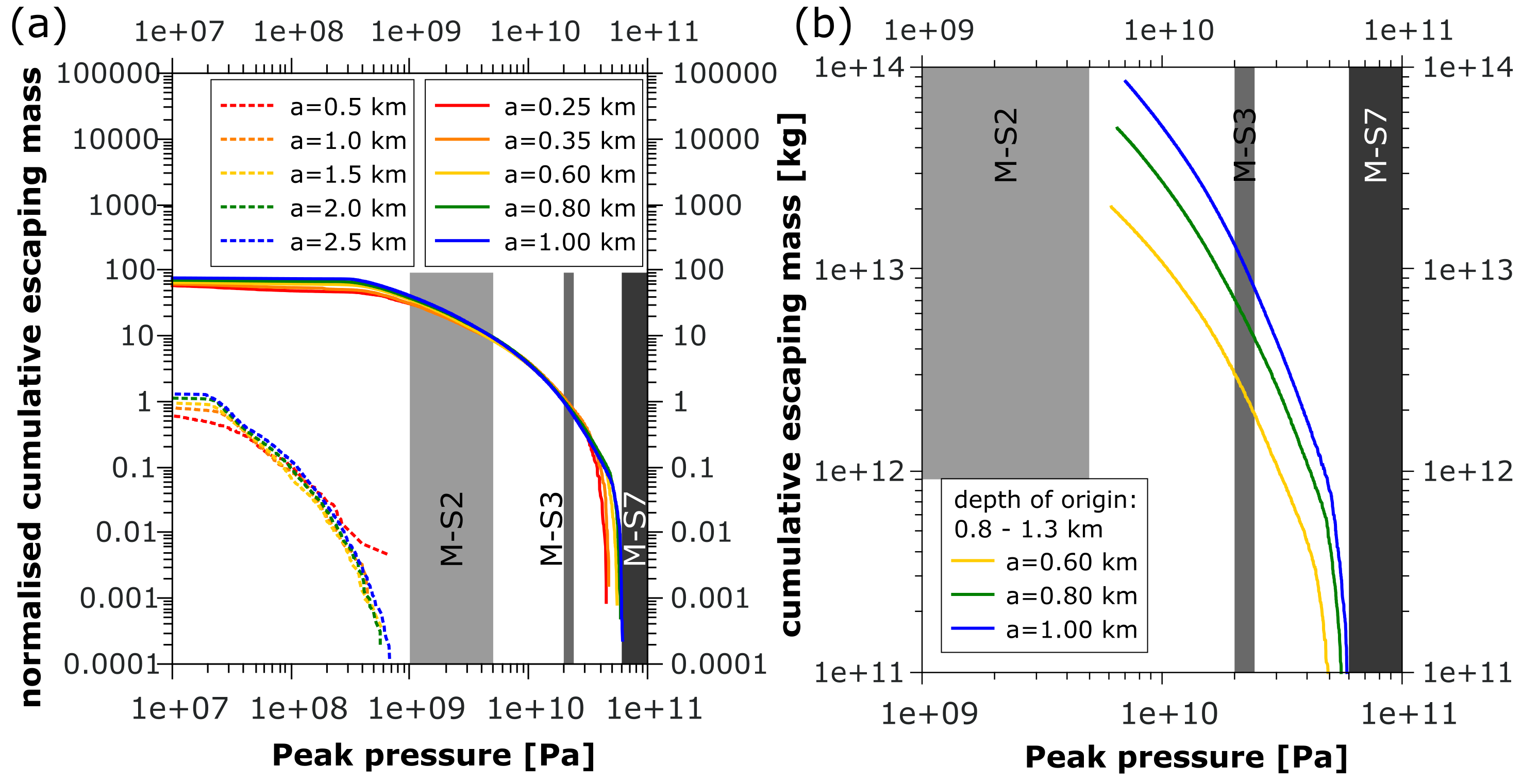}}
\caption{Cumulative mass per shock stage. It is shown the normalized escaping ejected material (a) and the fraction of this material, which originates from the formation depth between $800$ m and $1300$ m (b). The mass in (a) is normalized to the projectile mass. In (a), the cases of an impact velocity of $500$ m s$^{-1}$ and $5000$ m s$^{-1}$ are indicated by dashed and solid lines, respectively. The differently shaded grey bars indicate shock stages M-S2, M-S3 and M-S7 according to \cite{Stoeffler2018}.}
\label{fig6}
\end{figure}

In a recent study, \cite{Sturtz2022} considered differentiation of planetesimals that undergo protracted accretion starting close to CAIs formation and continuing to up to $\approx 2$ Ma. This accretion timescale is consistent with our result, since for the accretion law used by \cite{Sturtz2022} planetesimals grow to a relative radius of $>0.8$  within $0.4$ Ma. However, the differentiation of the parent body we obtained is also earlier than suggested by \cite{Sturtz2022}. This is because \cite{Sturtz2022} considered only the Al-Mg closure time of $2.255$ Ma calculated by \cite{Barrat2021} and considered it specifically as time of crystallization that occurred right after the source melt formation. However, recalculating the Al-Mg age for a $^{26}$Al half-life of $0.717$ Ma, \cite{Reger2023} showed that this closure time should be $2.14$ Ma and provided, in addition, another much earlier Al-Mg closure time of $1.83$ Ma. Here, we fitted U-Pb-Pb and Ar-Ar data that were not available to \cite{Sturtz2022} and included both Al-Mg ages for the accretion time discussion. A synthesis of these input data results in an early parent body differentiation. Another result from \cite{Sturtz2022} was a parent body radius of $70-130$ km. This matches neither our results of $R=20-30$ km for case (a), nor of $R\geq 170$ km for case (b). Also here, we clearly attribute this difference to our usage of a fit procedure that provides more precise statements.
\\
We obtained cooling rates of $140$ K to $170$ degrees per 1 Ma (cp. Fig. \ref{fig1}, right panels) within temperature intervals bracketed by $400$ K and $1470$ K. These cooling rates are slower by orders of magnitude than those of 5 degrees per 1 year and 1 degree per day suggested by \cite{Barrat2021} for different temperature intervals. Our cooling rates are, in fact, close to a maximum of $465$ degrees per 1 Ma suggested by \cite{Reger2023} for EC 002. As a local statement, in the time interval between both Al-Mg closure times our models produce cooling rates of $227$ K to $246$ K per 1 Ma, even closer to the estimate by \cite{Reger2023}. The modeled cooling rate within the closure temperature of the Ar-Ar system of $550\pm20$ K is $123$ K (case (a)) or $30$ K (case (b)), see Fig. \ref{fig1}, right panels. However, lack of $^{40}$Ar* diffusion \cite{Takenouchi2021} indicates a very fast cooling consistent with impact excavation. Thus, after cooling below the Ar-Ar closure temperature that occurs prior to the Ar-Ar closure time from \cite{Takenouchi2021}, the material at EC 002 layering depth within the parent body was reheated by an impact at $t\geq18$ Ma after CAIs and cooled again much faster after this excavation event.

\begin{deluxetable}{c c c c c c c}
\tabletypesize{\scriptsize}
\tablewidth{0.9\columnwidth}
\tablecaption{Escaping material for $v_{0}=5$ km s$^{-1}$.}
\centering

\tablehead{ & \multicolumn{3}{c}{\makecell{Escaping mass [kg]\\(fraction of total mass\\in brackets)}} & \multicolumn{3}{c}{\makecell{Escaping mass [kg] from depth of\\0.8 km - 1.3 km (fraction of total\\mass in brackets)}} }
\startdata
\makecell{Projectile\\diameter [km]} & total & $<10$ GPa & $<20$ GPa & total & $<10$ GPa & $<20$ GPa \\
\\ [-2.0ex]
1.2 & \makecell{$1.96\times 10^{14}$\\ } & \makecell{$1.86\times 10^{14}$\\(94.5 \%)} & \makecell{$1.93\times 10^{14}$\\(98.5 \%)} & \makecell{$2.03\times 10^{13}$\\ } & \makecell{$9.48\times 10^{12}$\\(46.7 \%)} & \makecell{$1.73\times 10^{13}$\\(85.1 \%)}\\ 
\\ [-2.0ex]
1.6 & $5.02\times 10^{14}$ & \makecell{$4.75\times 10^{14}$\\94.6 \%} & \makecell{$4.95\times 10^{14}$\\(98.6 \%)} & \makecell{$4.99\times 10^{13}$\\ } & \makecell{$2.28\times 10^{13}$\\(45.7 \%)} & \makecell{$4.27\times 10^{13}$\\(85.7 \%)}\\
\\ [-2.0ex]
2.0 & $1.06\times 10^{14}$ & \makecell{$1.01\times 10^{15}$\\(95.2 \%)} & \makecell{$1.05\times 10^{15}$\\(98.7 \%)} & \makecell{$8.50\times 10^{13}$\\ } & \makecell{$3.41\times 10^{13}$\\(40.1 \%)} & \makecell{$1.33\times 10^{13}$\\(84.4 \%)} 
\enddata
\label{table3}
\end{deluxetable}

The metallic core underwent thermal convection of a timescale of $<4$ Ma or $80$ Ma. It is unlikely that a core convection phase of $<4$ Ma would suffice to magnetize the chondritic part of the parent body or the kamacite fraction of EC 002. The core convection duration of up to $80$ Ma in case (b) is more likely to result in some magnetization, but we consider parent bodies from case (b) as unlikely due to a worse fit quality. In fact, \cite{Maurel2022} atributed a natural remnant magnetization in the kamacite fraction to the influence of the early solar nebula field. At the time of parent body accretion of $0.1$ Ma the solar nebula had not dissipated yet, allowing for such an influence.

The lower compacted part of the crust has a higher density than the mantle with a density contrast $300-700$ kg m$^{-3}$ (Fig. \ref{fig2}, left panel). However, this contrast does not suffice to induce subduction and mixing of this low-temperature region with the mantle due to its high viscosity. EC 002 originates from a partially (though almost fully) differentiated and transiently partially molten layer between this region and the fully differentiated mantle. A distance of $1.5-1.8$ km between the layering depth and the upper boundary of the magma ocean at its maximum extension (Fig. \ref{fig2}, right panel) implies that magma ocean melts likely were not involved in the genesis of the EC 002 material (Fig. \ref{fig2}). A magma ocean is, in general, characterized by a melt fraction of $>50$ \%, while EC 002 probably formed from an $\approx25$ \% melting of a nearly chondritic protolith. Since temperature curves fit the U-Pb-Pb data and the melting and metamorphic temperature constraints well at one and the same depth, and, in addition, are in a broad agreement with closure temperatures suggested for pyroxene and anorthite (Fig. \ref{fig1}), our models provide a more simple and straight-forward formation scenario for EC 002 without involvement of large-scale migration or extrusion of the source melt from a deeper region into a cold crust.

Ejection of material from a shallow depth of $<1$ km requires a less energetic meteorite-producing impact than an ejection from a deeper interior. The impact site would be covered with the rubble mixed from the undifferentiated crust, some impactor material, and some partially differentiated ejecta. Provided that the undifferentiated crust has a nearly chondritic composition and appears spectrally as such, the parent body may still be present in the solar system, but its spectrum would not match the spectral features of Erg Chech 002.

Depending on the impact velocity, the minimum projectile size to excavate and eject material faster than the escape velocity is $0.7-5$ km. Larger projectiles can excavate more material from the relevant depth. For most of this material, shock pressures remain below $20-40$ GPa and $1$ GPa for $5$ km s$^{-1}$ and $0.5$ km s$^{-1}$ impact velocity, respectively. This is below pressures to cause whole rock shock melting of the material\cite{Stoeffler2018}. Nevertheless, $1-5$ GPa are sufficient to induce shock deformations like fractures or mechanical twinning of pyroxene, and $\approx 20$ GPa are  sufficient for the formation of planar deformation features in plagioclase and partial conversion to diaplectic glass\cite{Stoeffler2018}. EC 002 is classified as weakly shocked (stage 1 IUGS 2007 / M-S2, see \cite{Barrat2021}), which corresponds to shock pressures of $1-5$ GPa. For the slow impact velocity of $500$ m s$^{-1}$, no material that is ejected and escapes the system was subject to such pressures (Fig. \ref{fig6} a). For the faster impacts, several shock stages can be found in the ejected material, and a large fraction of this material was shocked to stage M-S2 (Fig. \ref{fig6} a). Little material was shocked to pressures of $10$ or $20$ GPa (Table 4). However, for EC 002 it is also important to consider the depth of origin. The material originating from $800$ m to $1300$ m depth (Fig. \ref{fig6} b) shows a similar trend as the total escaping ejecta (Fig. \ref{fig6} a), and this material was subject to pressures of $\approx 6$ GPa or more. This is just above the estimated pressure range of the shock stage M-S2. About $40-50$ \% of the escaping material from the estimated formation depth was shocked to $<10$ GPa, and about $85$ \% of the material was shocked below a pressure of $20$ GPa, corresponding to shock stage M-S3 (Table 4). Consequently, an impact with a velocity of $5$ km s$^{-1}$ for projectiles larger than $700$ m can eject material like EC 002 from its formation depths at sufficiently high ejection velocities to leave the parental system. Impacts with a velocity of $500$ m s$^{-1}$ could eject the same material, but pressures are not sufficient to shock the material to stage M-S2. For faster impacts than $500$ m s$^{-1}$, the shock pressure increases, so that we conclude that the minimum impact event has a velocity of more than $500$ m s$^{-1}$ and an impact energy of $\approx 2-3 \times 10^{19}$ J. When accounting for the effect of impact angle, the required impact energy could be even larger compared to our vertical impact simulations, because the excavation depths decrease with shallower impacts. Further, also large-scale impacts that destroy large parts of the parental body could eject material like EC02. However, due to the size frequency of the asteroid reservoir, we argue that such big impacts are much less probable compared to our proposed scenario.

\begin{acknowledgments}
This work was supported by the Deutsche Forschungsgemeinschaft (DFG) [project number 434933764] and by the Klaus Tschira Foundation.
\end{acknowledgments}

%



\appendix

\section{Model Description} 
This study uses a 1D finite differences thermal evolution model for planetesimals heated mainly by $^{26}$Al similar to those in \cite{Neumann2018b} and \cite{Neumann2021} that calculates thermal evolution of small porous planetary bodies, their compaction from an initially unconsolidated state due to hot pressing, and metal-silicate separation, by solving a number of equations that describe these processes. The model description is provided in the following.

A non-stationary 1D heat conduction equation in spherical coordinates is solved for the temperature by the finite differences method along the spatial and temporal domain:
\begin{eqnarray}
\rho c_{p} \left( 1+x_{Fe}S_{Fe}+x_{Si}S_{Si}\right) \frac{\partial T}{\partial t} =\frac{1}{r^{2}}\frac{\partial}{\partial r}\left( k r^{2} \frac{\partial T}{\partial r} \right) + Q(r,t) \text{,}
\end{eqnarray}
with the bulk density $\rho$, the heat capacity $c_{p}$, the initial fractions $x$ and Stefan numbers $S$ for metal (Fe) and silicates (Si, see Table \ref{tablea1}), the temperature $T$, the time $t$, the radius variable $r$, and the energy source density $Q$. See \cite{Neumann2018b} for the details on Stefan number. The energy source for the heating is the radioactive decay of the isotopes $^{26}$Al, $^{60}$Fe, $^{40}$K, $^{232}$Th, $^{235}$U, and $^{238}$U that is homogeneous throughout the planetesimal depth prior to the differentiation:
\begin{eqnarray}
Q(r,t) = \rho \sum_{i} f_{i}Z_{i}\frac{E_{i}}{\tau_{i}}\exp \left( -\frac{t-t_{0}}{\tau_{i}} \right) \text{,} \label{heso}
\end{eqnarray}
with the bulk density $\rho$, the number of atoms of a stable isotope per 1 kg of the primordial material $f$, the initial ratio of radioactive and stable isotope $Z$, the decay energy $E$, the mean life $\tau=\lambda/\log(2)$, the half-life $\lambda$, and the accretion time $t_{0}$ of the planetesimal (Table \ref{tablea2}). The radionuclides are distributed homogeneously at the time $t_{0}$, but their distribution is prone to variation with the porosity $\phi$. The porosity is initially constant throughout the planetesimal. It develops inhomogeneously with depth during the thermal evolution under the action of temperature and pressure. Further inhomogeneitites arise later on when metal and silicates separate. In this case, the summands in Eq. (\ref{heso}) are scaled with the local ratio of the current volume fraction to the initial volume fraction of metal for $^{60}$Fe or of silicates for all other radionuclides.

\subsection{Fitting Procedure}
An approach utilized in several H and L chondrite parent body studies \cite[e.g.,][]{Henke2012,Gail2019} and a study of the Acapulco-Lodran parent body \cite{Neumann2018b} is used to fit the thermo-chronological data with a least square procedure. The fit procedure provides those initial model parameters that result in a thermal evolution which reproduces the thermo-chronological data as close as possible. For this, thermal evolution models are calculated for a given set of initial model parameters. All data derives from the single meteorite EC002 and, thus, to one single depth within a parent body. First, the fit quality at any depth is determined by calculating the distances of the temperature curves $T(t,d)$ to each of the data points.

We denote the measured closure time and the closure temperature of one of the radioactive decay systems (enumerated with index $i$) as $t_{i}^{c}$ and $T_{i}^{c}$, respectively. The temperature curve $T(t,d)$ has in general a shape with an increase on a time scale of $<1$ Ma to several tens of million years until the maximum temperature $T_{max}$ is achieved at a certain time $t_{max}$, followed by a cooling phase. To determine the quality of the model, we distinguish between two cases for each data point:

1. If the maximum temperature $T_{max}$ is higher than the corresponding closure temperature $T_{i}^{c}$, the summed distance of the temperature curve to the data points is defined by
\begin{eqnarray}
\delta^{2} (d)=\sum_{i} \left( \frac{\left( t_{i}^{c}-t\left( T_{i}^{c} \right) \right)^{2}}{\sigma_{t,i}^{2}}+\frac{\left( T_{i}^{c}-T\left( t_{i}^{c},d \right) \right)^{2}}{\sigma_{T,i}^{2}} \right) ,
\end{eqnarray}
where $t\left( T_{i}^{c} \right)$ is the time at which the temperature curve passed the closure temperature $T_{i}^{c}$ on its descending branch, $T\left( t_{i}^{c},d \right)$ is the temperature achieved at the depth $d$ at the closure time $t^{c}$, and $\sigma_{t,i}$ and $\sigma_{T,i}$ are the errors of the determination of the closure temperatures and cooling ages, respectively.

\begin{table*}
\centering
\caption{Parameters used in the models. Note that the mass and volume fractions represent initial values, while local values at time $t$ may change due to differentiation. The volume fraction of the phase $i$ is obtained from $v_{i}=x_{i}\rho_{g}/\rho_{i}$.}
\centering
\begin{tabular}{lrrrr}
\hline \\ [-1.7ex]
Variable & Symbol & Unit & Value \\
\\ [-1.7ex]
\hline
\\ [-1.4ex]
Ambient temperature & $T_{\text{S}}$ & K & $290$ \\
\\ [-1.4ex]
Initial porosity & $\phi_{0}$ & - & $0.5$ \\
 \\ [-1.4ex]
Matrix grain size & $b_{m}$ & m & $10^{-4}$ \\
\\ [-1.4ex]
Effective stress & $\sigma$ & Pa & see text and \cite{Neumann2014} \\
\\ [-1.4ex]
Gas constant & $\mathcal{R}$ & J mol$^{-1}$K$^{-1}$ & $8.314472$ \\
\\ [-1.4ex]
Metal th. conductivity & $k_{Fe}$ & W m$^{-1}$K$^{-1}$ & $10$ \\
\\ [-1.4ex]
Silicate th. conductivity & $k_{Si}$ & W m$^{-1}$K$^{-1}$ & $4.3$ \\
\\ [-1.4ex]
Grain density & $\rho_{g}$ & kg m$^{-3}$ & $3690$ \\
\\ [-1.4ex]
Metal density & $\rho_{Fe}$ & kg m$^{-3}$ & $6811$ \\
\\ [-1.4ex]
Silicate density & $\rho_{Si}$ & kg m$^{-3}$ & $3270$ \\
\\ [-1.4ex]
Metal mass fraction & $x_{Fe}$ & - & $0.219$ \\
\\ [-1.4ex]
Silicate mass fraction & $x_{Si}$ & - & $0.781$ \\
\\ [-1.4ex]
Metal vol. fraction & $v_{Fe}$ & - & $0.12$ \\
\\ [-1.4ex]
Silicate vol. fraction & $v_{Si}$ & - & $0.88$ \\
\\ [-1.4ex]
Metal solidus & $T_{Fe,S}$ & K & $1213$ \\
\\ [-1.4ex]
Metal liquidus & $T_{Fe,L}$ & K & $1700$ \\
\\ [-1.4ex]
Silicate solidus & $T_{Si,S}$ & K & $1425$ \\
\\ [-1.4ex]
Silicate liquidus & $T_{Si,L}$ & K & $1850$ \\
[0.6ex]
\hline
\end{tabular}
\\ [2ex]
\label{tablea1}
\end{table*}

2. If the maximum temperature $T_{max}$ does not surpass the closure temperature $T_{i}^{c}$ throughout the model time, the distance is determined by
\begin{eqnarray}
\delta^{2} (d)=\sum_{i} \left( \frac{\left( t_{i}^{c}-t_{max} \right)^{2}}{\sigma_{t,i}^{2}}+\frac{\left( T_{i}^{c}-T_{max} \right)^{2}}{\sigma_{T,i}^{2}} \right).
\end{eqnarray}

Metamorphic or melting temperatures appropriate for the meteorite involved can be utilized for defining penalty functions to penalize the depths at which the maximum temperature does not agree with the metamorphic constraints:
\begin{eqnarray}
\mathcal{P}=\max \lbrace T_{max}-T_{u},0\rbrace-\min\lbrace T_{max}-T_{l},0\rbrace ,
\end{eqnarray}
where $T_{u}$ is the least upper and $T_{l}$ is the highest lower bound on the metamorphic temperature. Penalty functions ensure that temperature curves with unrealistically high maxima that contradict meteorite metamorphic temperature ranges but still would fit the data well on their descending branches are excluded. The values of $\mathcal{P}$ are zero within the allowed temperature ranges and add a rapidly growing penalty the more the maximum $T_{max}$ of a temperature curve deviates from this range. For each depth $d$, the value of $\mathcal{P}$ is added to the value of $\delta^{2}(d)$: $\tilde{\delta}^{2}(d)=\delta^{2}(d)+\mathcal{P}$.

With $\tilde{\delta}^{2}=\min_{d}\tilde{\delta}^{2}(d)$, we define the normalized quality function:
\begin{eqnarray}
\chi_{n}=\left[\frac{1}{n}\tilde{\delta}^{2} \right]^{\frac{1}{2}}
\end{eqnarray}
by which we judge how good a given planetesimal fits our data set (here, $n$ denotes the total number of the data points). The depth at which the minimum $\tilde{\delta}^{2}$ is attained defines the layering depth of EC002.

The data set comprises in total three data points with available closure times and temperatures from U-Pb-Pb chronometer for pyroxene and phosphates and Ar-Ar chronometer for plagioclase, i.e., $n=3$. Thus, $\delta^{2} (d)$ has either two or three summands depending on whether the Ar-Ar data is excluded (case (a)), or included (case (b)). 
To account for the melting degree of the source melt suggested by \cite{Barrat2021}, we added a penalty function $\mathcal{P}$ assuming with $T_{l}=1460$ K and $T_{u}=1520$ K.
For each case (a) or (b), this procedure results in one layer at the layering depth of EC002 for each planetesimal considered and in a best-fit planetesimal with the smallest fit quality.

\begin{table*}
\centering
\caption{Parameters used for the calculation of radiogenic energy. The element mass fractions refer to stable isotopes, the initial ratios are between unstable and stable isotopes of an element, and the decay energies are per particle. The number of atoms of the stable isotope per $1$ kg of the primordial material is $f=xN_{\text{A}}/m_{\text{a}}$ with the relative mass fraction $x$ of the stable isotope, the molar mass of the radioactive isotope $m_{\text{a}}$ in kg, and the Avogadro number $N_{\text{A}}$. All values are referenced in \cite{Neumann2018b}.}
\centering
\begin{tabular}{c|llllll}
\hline \\ [-1.7ex]
Isotope & \multicolumn{1}{c}{$^{26}$Al} & \multicolumn{1}{c}{$^{60}$Fe} & \multicolumn{1}{c}{$^{40}$K} & \multicolumn{1}{c}{$^{232}$Th} & \multicolumn{1}{c}{$^{235}$U} & \multicolumn{1}{c}{$^{238}$U} \\
\\ [-1.7ex]
\hline
\\ [-1.4ex]
Element mass fr. $x$ & $1.21\cdot 10^{-2}$	& $2.41\cdot 10^{-1}$ &	$7.07\cdot 10^{-4}$	& $5.16\cdot 10^{-8}$ & $2.86\cdot 10^{-8}$ & $2.86\cdot 10^{-8}$ \\
\\ [-1.4ex]
Half-life $\lambda$ [years] & $7.17\cdot 10^{5}$ & $2.62\cdot 10^{6}$ & $1.25\cdot 10^{9}$ & $1.41\cdot 10^{10}$ & $7.04\cdot 10^{8}$ & $4.47\cdot 10^{9}$ \\
\\ [-1.4ex]
Initial ratio $Z$ & $5.25\cdot 10^{-5}$ & $1.15\cdot 10^{-8}$ & $1.50\cdot 10^{-3}$ & $1.0$ & $0.24$ & $0.76$ \\
\\ [-1.4ex]
Decay energy $E$ [J] & $4.99\cdot 10^{-13}$ & $4.34\cdot 10^{-13}$ & $1.11\cdot 10^{-13}$ & $6.47\cdot 10^{-12}$ & $7.11\cdot 10^{-12}$ & $7.61 \cdot 10^{-12}$ \\
[0.6ex]
\hline
\end{tabular}
\\ [2ex]
\label{tablea2}
\end{table*}

\subsection{Porosity}
The evolution of the bulk pore space volume fraction, i.e., porosity $\phi$ is calculated using a time-dependent differential equation that establishes a relation between the strain rate $\dot{\varepsilon}$ and the applied stress $\sigma$. The average local porosity $\phi$ is obtained from the average local strain rate $\dot{\varepsilon}$ that is described with an Arrhenius term derived for the diffusion creep \cite{Schwenn1978}:
\begin{eqnarray}
\frac{\partial \log ( 1-\phi )}{\partial t} &=& \dot \varepsilon \nonumber \\
&=& 1.26\cdot 10^{-18} \sigma^{1.5} b^{-3} \exp \left(-\frac{356}{\mathcal{R}T} \right) \text{,} \label{cr2}
\end{eqnarray}
Here, the stress $\sigma$ is in Pa, the grain size $b$ in m, the activation energy $\mathcal{E}$ in kJ mol$^{-1}$, the gas constant $\mathcal{R}$ in kJ and the temperature $T$ in K. Thereby, the compaction behavior of the precursor material is assumed to be similar to that of ordinary chondrites and is approximated with the olivine diffusion creep.


A grain size of $1-1.5$ mm observed for EC 002 \cite{Barrat2021} is a consequence of grain growth and melt crystallization. It does not represent the grain size $b$ of the dust grains the parent body accreted from that is required for the calculation of compaction. Typical grain sizes of undifferentiated porous ordinary chondrites range from $10^{-6}$ m (fine-grained matrix, $\mu$m-scale) to $10^{-3}$ m (chondrules, mm-scale). As a compromise, we use a value of $b=10^{-4}$ m. An initial porosity of $\phi_{0}=0.5$ is a typical value based on the porosities of the random loose and random close packings \cite[e.g.,][]{Henke2012,Neumann2014}. The effective stress $\sigma$ is calculated for a specific ordered packing of equally sized spheres \cite{Neumann2014}. Here, we use the simple cubic packing that has a porosity of $\approx 50$ \%.

\subsection{Melting and Metal-Silicate Differentiation}
The model routine for melting and separation of metal and silicates is based on our model for the parent body of Acapulco-Lodran meteorites \cite{Neumann2018b}. We refer to this paper for details. The silicate melt is produced between a solidus and a liquidus temperature (see Table \ref{tablea1}), where the melt fraction as a function of temperature is calculated according to \cite{McKenzie1988}. The melting of metal is assumed to occur linearly between the metal solidus and liquidus temperatures. The separation of metal from the silicate rock is modeled by considering percolation of metal (with a viscosity of $10^{-2}$ Pa s) or silicate melts (a viscosity of $10^{3}$ Pa s) in a partially molten system and by considering Stokes settling of metal in a magma ocean. Here, we consider a grain size with an initial diameter of $2\cdot 10^{-6}$ m that grow to a grain size with a diameter of $10^{-3}$ m observed for EC002. Provided a sufficiently high melt velocity, local fractions of metal melt are relocated towards the center, while the partial silicate melt is relocated towards the surface, based on the respective velocity computed. The volume occupied previously by the melt is compensated by the upward (for the migration of metallic melt) or downward (for the migration of silicate melt) matrix compaction. As a result, a metallic core and a silicate mantle can form. Both layers can be largely molten, thus, a magma ocean may form in the mantle and both core and mantle could experience transient convection periods.

\subsection{Material Properties}
The mechanical and thermal material properties are weighted averages of those of the phases and of the pore space:
\begin{eqnarray}
&&\rho=(1-\phi)\rho_{g}=(1-\phi)\left( v_{Fe}\rho_{Fe}+v_{Si}\rho_{Si} \right) \text{,}\\
&&k={k_{Fe}}^{v_{Fe}}{k_{Si}}^{v_{Si}}\left( e ^{-4\phi / \phi_{1}} + e^{-4.4-4\phi / \phi_{2}} \right)^{1/4} \\
&&c_{p}=x_{Fe}c_{p,Fe}+x_{Si}c_{p,Si} \text{,}
\end{eqnarray}
with the porosity $\phi$, the thermal conductivities $k_{i}$, volume fractions $v_{i}$, mass fractions $x_{i}$, and specific heats $c_{p,i}$ of the phases, as well as constants $\phi_{1}=0.08$ and $\phi_{2}=0.17$. The grain density of the compacted material with $\phi = 0$, $\rho_{g}=3690$ kg m$^{-3}$ is an average value calculated for the ordinary chondritic material with volume fractions and metal and silicate densities as in Table \ref{tablea1}. Note that both mass and volume fractions change during the differentiation accordingly. For all other parameters see either Table \ref{tablea1}, or \cite{Neumann2018b}.

\subsection{Magma Ocean}
While the source melt fraction estimate for EC002 does not indicate a high melt fraction associated with a magma ocean, it is thinkable that more melt was produced in a deeper interior of the parent body. Thus, the model considers formation of a magma ocean and its cooling due to the liduid-state convection that also influences cooling of the parent body as a whole. A magma ocean is a mantle region where the temperature is high enough to produce more than $50$ \% melt. For an ordinary chondritic composition, this corresponds to a temperature of $\gtrsim 1650$ K. Above $T=1650$ K, the thermal conductivity is substituted with the effective thermal conductivity $k_{eff}$, that simulates cooling by convection in a mixed iron-silicate magma ocean:
\begin{eqnarray}
k_{eff}=0.089 kRa^{1/3} \text{,}
\end{eqnarray}
where $Ra$ is the Rayleigh number \cite[see, e.g.,][]{Neumann2018a}. A magma ocean can, in principle, form in a certain range of parameter value combinations with respect to parent body accretion time and size. For planetary objects with a size of $<1000$ km, it is only transient and solidifies on a time scale of a few hundred million years or less.

\begin{figure}
\begin{minipage}[ht]{8cm}
\setlength{\fboxsep}{0mm}
\centerline{\includegraphics[trim = 10mm 0mm 0mm 0mm, clip, height=6cm]{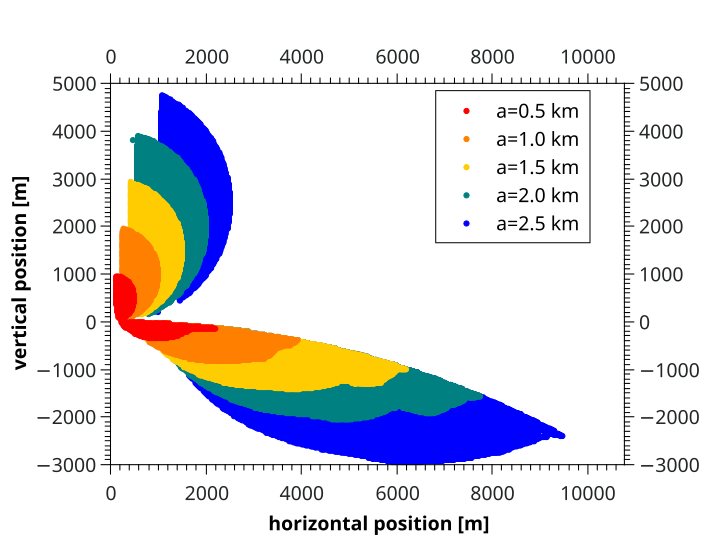}}
\end{minipage}
\begin{minipage}[ht]{8cm}
\setlength{\fboxsep}{0mm}
\centerline{\includegraphics[trim = 10mm 0mm 0mm 0mm, clip, height=6.3cm]{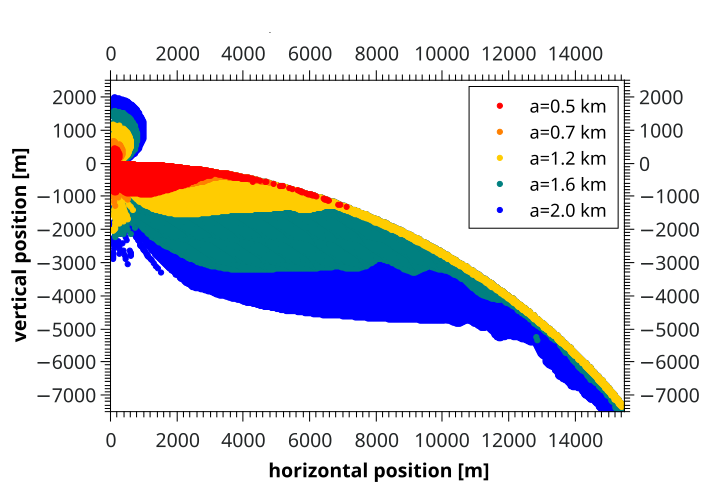}}
\end{minipage}
\caption{Initial position of ejected projectile and target material for an impact velocity of 500 m s$^{-1}$ (top panel) and 5000 m s$^{-1}$ (bottom panel).}
\label{figa1}
\end{figure}

\subsection{Radius Change}
The radius $\overline{R}(t)$ of the object considered changes with the bulk porosity $\phi_{bulk}(t)$ of the entire object at the time $t$, obtained by integrating the local porosity over the radius $r$, according to 
\begin{eqnarray}
\overline{R}(t)=(1-\phi_{bulk}(t))^{-1/3}R \text{,} \label{radius}
\end{eqnarray}
where $R$ is the reference radius, i.e., the radius that would be attained if the porosity were zero. For better understanding, we associate parent bodies with their reference radius $R$ in the main manuscript. It is used for the analysis of the results since it is representative for sets of bodies with equal mass and grain density, but different porosity. The relation $\overline{R}(t)>R$ is always true and $\overline{R}(t)=R$ would be only possible for $\phi_{bulk}(t)=0$, which never occurs in the calculations.

All equations involved are solved on the spatial radius domain ranging from the center of the planetesimal up to its surface. The spatial grid is transformed from $0\leq r\leq R$, with the distance from the center $r$ in m, to $0\leq \eta \leq 1$ using the transformation $\eta :=r/\overline{R}(t)$. The time and space derivatives are transformed as well and the transformed expressions are applied to all equations involved, such that features like Langrangian transport of porosity \cite[][Eq. (10)]{Neumann2012} and other quantities are accounted for. While the positions of the grid points between $0$ and $1$ are fixed, the variable values at the grid points are updated at every time step according to the above transformations. A number of grid points is chosen in such way that the distance between the grid points in the interior is $\approx 100$ m, while in the outer $3$ km where strong temperature and porosity gradients occur it is $\approx 10$ m. Non-stationary equations are discretized also with respect to the time variable $t$ and solved using implicit finite difference method.

\bibliography{main}{}
\bibliographystyle{aasjournal}

\end{document}